\documentstyle[aas2pp4, psfig]{article}

\begin{document}
 
\title{Lensing Effects on the Protogalaxy Candidate
cB58 and their Implications for the Cosmological Constant}
\author{T.Hamana, M.Hattori\altaffilmark{1}}
\affil{Astronomical Institute, Tohoku University, Sendai 980-77, Japan}
\author{H.Ebeling, J.P.Henry}
\affil{Institute for Astronomy, University of Hawaii, 2680 Woodlawn Dr,
Honolulu, HI 96822, USA}
\author{T.Futamase and Y.Shioya}
\affil{Astronomical Institute, Tohoku University, Sendai 980-77, Japan}
  
\altaffiltext{1}{also at the Max-Planck-Institut f\"{u}r
Extraterrestrische Physik (MPE), D-85740 Garching, Germany} 
 

\begin{abstract} 
The amplification of the protogalaxy candidate cB58 due to
gravitational lensing by the foreground cluster of galaxies
MS1512.4+3647 is quantified based on recent ROSAT and ASCA X-ray
observations.  It is found that the amplification is at most 25 for
any reasonable cosmological model with or without cosmological
constant. It is also argued that the system may be used to place new
constraints on the value of the cosmological constant. The gas
mass fraction for this cluster is found to be about 0.2.
\end{abstract} 


\section{Introduction}
The detection of galaxies at their formation epoch has been a
long-standing goal. Yee et al.\ (1996) recently reported the discovery
of a protogalaxy candidate, designated cB58, at a high redshift of
$z=2.72$.  Ellingson et al.\ (1996) concluded that the spectral energy
distribution of cB58 is most consistent with a population synthesis
model (Bruzual \& Charlot 1993) for young age ($\sim$ a few $10^7$ yr)
and reddening of $E(B-V) \sim 0.3$.  The star formation rate (SFR)
inferred from the system's extremely high luminosity of ${\rm M_v\sim
-26}$ is 4700 ${\rm M_{\odot}yr^{-1}}$ ($H_0=75$ km s$^{-1}$
Mpc$^{-1}$, q$_0=0.1)$.  Such a high SFR is enough to make all the
stars in a giant galaxy within a few $10^8$ years.  This indicates
that cB58 may be the long-sought galaxy in the initial starburst
phase.  When corrected to zero extinction the SFR value is still as
high as 400 ${\rm M_{\odot} yr^{-1}}$.  These values are extremely
high compared to a mean SFR of 9.3 ${\rm M_{\odot}yr^{-1}}$ for other
distant galaxies (Steidel et al. 1996a) where zero extinction is
assumed.  
The central surface brightness of cB58 is about 21 mag/arcsec$^2$ in the V 
band (Yee et al. 1996). This is about 2 mag brighter that the mean central 
surface brightness of high-$z$ ($z \sim 3$) star forming galaxies but
comparable
to the brightest known examples (Giavalisco et al. 1996, Steidel et al. 
1996b). Thus cB58 could be regarded as one of the most active star-forming
galaxy known.

However, one should consider the possible effect of amplification by
gravitational lensing because, in projection, cB58 lies close ($\sim
6''$) to the center of the cluster of galaxies MS1512.4+3647 which is
in the foreground at $z=0.373$.  Williams and Lewis (1996) claimed
that the phenomenal properties of this protogalaxy candidate are in
fact due to amplification by gravitational lensing by the foreground
cluster, and that the unlensed properties of the source are typical of
high-redshift starforming systems.  The 
high surface brightness of cB58 does not rule out the lensing hypothesis,
since lensing conserves surface brightness, but does require that it be
atypical. 
However, the lensing amplification
is rather sensitive to the assumed cluster mass distribution model.
Therefore, detailed knowledge of the cluster mass distribution is very
important to quantify the impact of amplification by lensing and to
judge whether the apparently huge star formation rate of cB58 is only
an artefact.  X-ray observations are ideal for quantifying the cluster
mass distribution.  Assuming hydrostatic equilibrium for the X-ray
emitting hot gas distribution, the measurement of the temperature and
density distribution of the hot intra-cluster gas give good
constraints on the cluster mass distribution. Here, we reexamine the
lensing amplification effect in cB58 due to the cluster MS1512.4+3647
by constraining its mass distribution using temperature data obtained
from ASCA and ROSAT PSPC observations and a gas distribution
determined from ROSAT HRI observations.

Finally, we argue that cB58 provides a new opportunity to place
constraints on the value of the cosmological constant, primarily
because there is no lensed counter image (see Fig. 2 of Yee et al.,
1996).

Throughout this paper, the Hubble constant $H_0$ is taken to be
75$h_{75}$ km s$^{-1}$ Mpc$^{-1}$.

 
\section{X-ray results for the cluster lens} 


\subsection{X-ray spectral analysis} 

\begin{figure}
\vskip6cm 
\special{epsfile=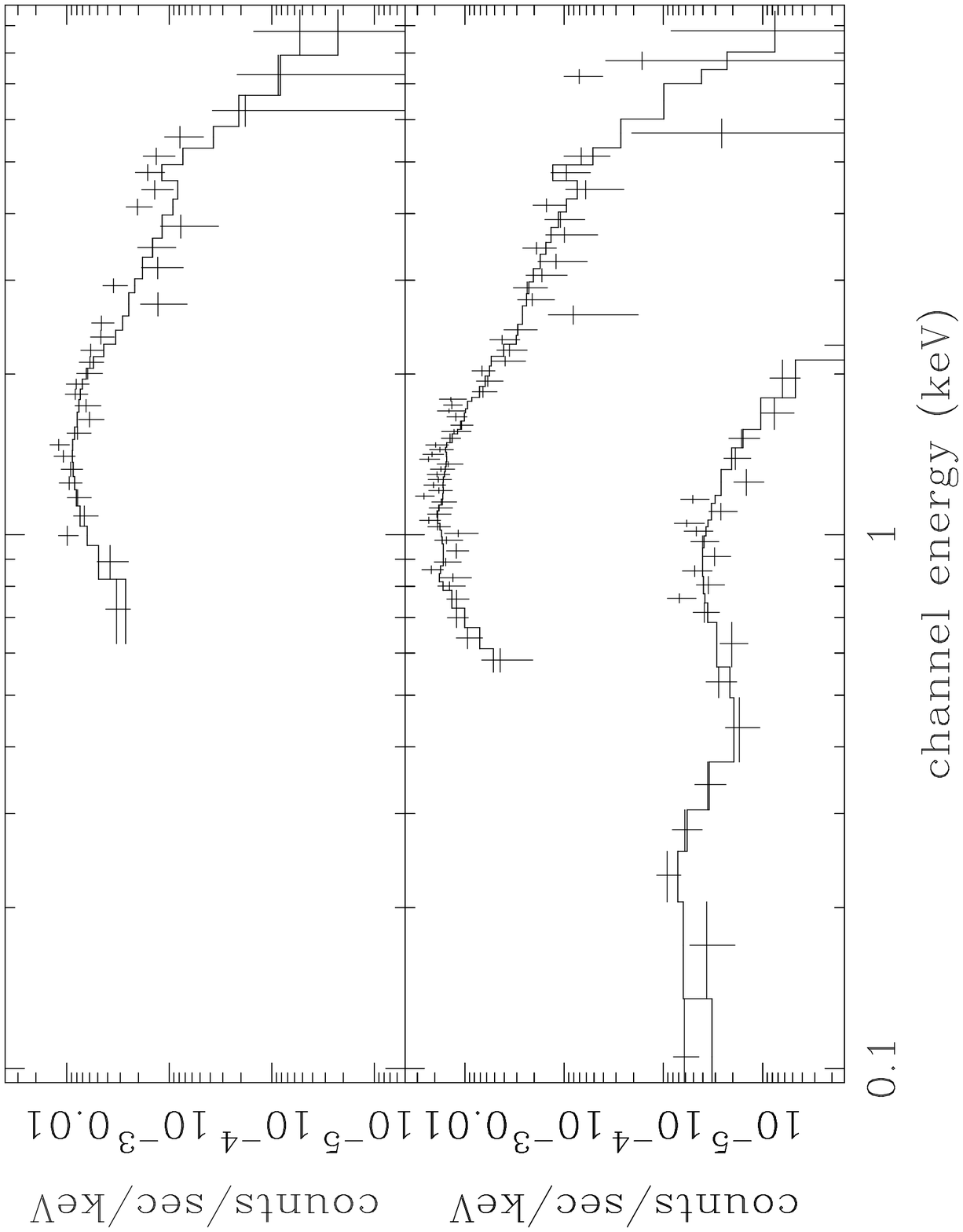 rotation=-90 hoffset=240 voffset=200
hscale=0.35 vscale=0.35}
\figcaption[Fig1.ps]{X-ray spectrum of MS1512.4+3647 obtained with the
ASCA GIS and SIS, as well as with the ROSAT PSPC.  The summed spectra
of GIS~2 and 3, and SIS~0 and 1, are shown in the upper and lower
panels, respectively.  The pulse height spectra are re-binned so as to
contain at least 40 photons for the GIS and 20 photons for the SIS
spectra in each bin. At lower energies, the lower panel also shows the
spectrum obtained with the ROSAT PSPC. The pulse height spectrum is
re-binned such that the count rate in each bin represents a signal to
noise ratio of at least 3.  A Raymond-Smith model which gives the best
fit for the GIS, SIS, and PSPC spectra simultaneously is
superposed.\label{fig1}} 
\end{figure}

ASCA observations of MS1512.4+3647 were performed in January 1995 with
an effective exposure time of 19 ks.  The SIS observations were
performed in 2 CCD Faint Mode for high bit rate observations and 2 CCD
Bright Mode for medium bit rate observations. The Faint Mode data were
converted to Bright Mode using the standard {\sc Ftools} software.
Additional data reduction was done with the standard {\sc Ftools} and
{\sc Xspec} packages. Figure 1 shows the resulting X-ray spectrum of
MS1512.  For GIS~2 and GIS~3, source counts were extracted from a
circle of 6 arcmin radius centered on the source; for the SIS, a
circle of 4.3 arcmin radius centered on the source was selected.  The
background was taken from the blank sky observation obtained at the
same detector positions. We use the summed spectra of the two GIS and
the two SIS for the spectral analysis. ROSAT PSPC observations of
MS1512.4+3647 were performed in August 1992 with an exposure time of
5.2 ks. The source spectrum was extracted from a 2.5 arcmin radius
circle centered on the cluster after the point source about $1'$ south
of the cluster had been masked out. The background was taken from a
4.16 arcmin wide annulus around the source region within which all
apparent sources had been masked out. The best fitting Raymond-Smith
parameters obtained in a simultaneous fit to the ASCA and ROSAT data
are summarized in Table 1 together with their 68\% confidence errors
for one interesting parameter, 
where the redshift is
fixed to that of MS1512.4+3647, $z=0.37$. 
The best-fit value for the hydrogen
column density is consistent with the Galactic value of $1.4 \times
10^{20}$ cm$^{-2}$ of Dickey and Lockman (1990). Figure 2 shows
$\chi^2$ contours for 68\%, 90\% and 99\% confidence for two
interesting parameters, temperature and $N_H$. The $N_H$ value is well
constrained by the ROSAT PSPC spectrum; the main source of error in
the temperature measurement are the photon statistics of the ASCA
data.

\begin{figure}
\vskip6cm
\special{epsfile=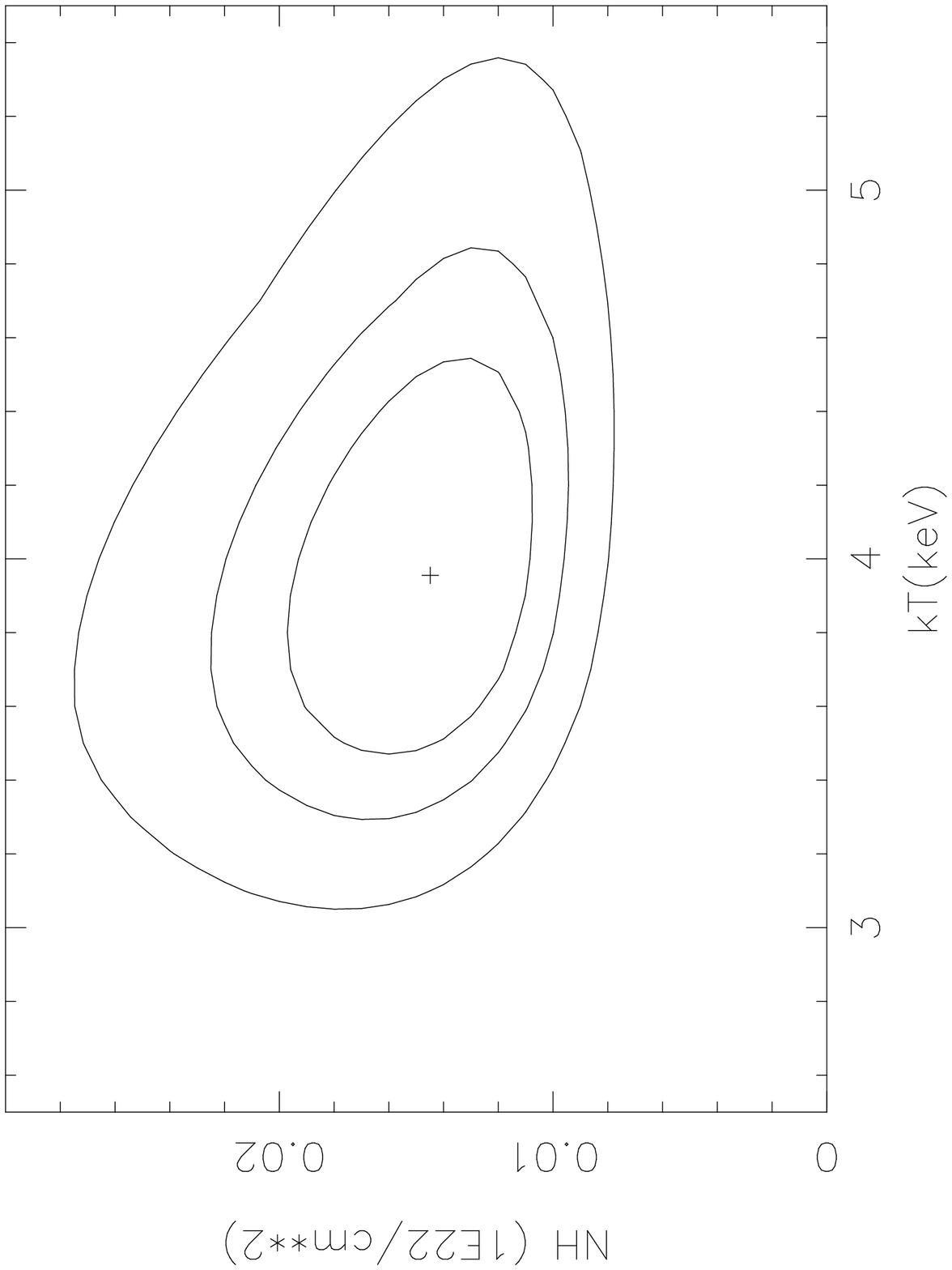 rotation=-90 hoffset=240 voffset=200
hscale=0.35 vscale=0.35}
\figcaption[Fig2.ps]{$\chi^2$ contours for 68\%, 90\% and 99\%
confidence obtained by spectral fitting of the ASCA and ROSAT
data. The plus sign marks the best fit values of column density and
temperature. \label{fig2} } 
\end{figure}

\begin{table*}
\caption{X-ray properties of MS1512.4+3647 from the spectral analysis of
 ASCA/ROSAT data.\label{tab1}}
\begin{tabular}{cccccc}
\hline\hline
$k_{\rm b}T_X$      &        $L_X$ ($2-10$ keV)              &    $Z_{\rm Fe}$     &      $N_{\rm H}$      & $\chi^2$/d.o.f. \\
   (keV)            & ($10^{44}$ h$_{75}^{-2}$ erg s$^{-1}$) &    $(Z_{\odot})$    & $(10^{20}$ cm$^{-2}$) &                 \\  
$3.9^{+0.5}_{-0.3}$ &           $1.7_{-0.1}^{+0.1}$          & $0.4_{-0.2}^{+0.2}$ & $1.5_{-0.3}^{+0.3}$   & 101.4/110$=$0.92\\
\hline 
\end{tabular}
\end{table*}


\subsection{X-ray spatial analysis} 
ROSAT HRI observations were performed in February 1995 with an
exposure time of 35 ks. Figure 3 shows the X-ray image of
MS1512.4+3647 in pulse height channels $1-7$ only since there is no
significant emission from MS1512.4+3647 in channels $8-15$.  The bin
size is 1$\times$1 arcsec$^2$.  The astrometry is based on the
position of the bright star south of MS1512.4+3647.  Since the X-ray
source at this position is possibly double -- the poor photon
statistics preclude a definitive statement -- the resulting astrometry
solution is only known to about 2 arcsec.  The image has been
adaptively smoothed using a Gaussian kernel whose size is determined
from the requirement that 16 photons be covered by the kernel (Ebeling
et al.\ 1995). The resulting Gaussian widths of the kernel range from
1.9 to 17.5 arcsec.  The peak of the emission is pronounced and
centered on the cluster cD.  The emission from the inner 30
h$_{75}^{-1}$ kpc radius accounts for some 15 per cent of the overall
cluster emission.  Although the very core region within the inner
$8''$ appears spherically symmetric, the emission soon becomes
significantly elongated at sizes well resolved with the HRI.  Given
the resolution of the HRI, there might be an asphericity in the
surface brightness even in the very central region of the cluster at
the (projected) location of cB58.

The radial surface brightness profile is shown in Figure 4.  The point 
source about $1'$ south of the cluster was again excluded. There is
some evidence for excess emission in the very core of the cluster
which could indicate the existence of a cooling flow. However, the
radial extent of the excess emission is less than 2 arcsec which is
somewhat less than the HRI resolution. Also the overall significance
of the excess over a pure $\beta$ model is less than $3\sigma$.
Therefore, the evidence for a cooling flow in this cluster is
marginal. We have fit a $\beta$ model to the azimuthally averaged
radial surface brightness $I(\theta)$. This model assumes
\begin{equation}
  I(\theta) = I_0 \left[ 1+\left(\frac{\theta}{\theta_c}\right)^2
\right]^{-3\beta+1/2}.
\end{equation}
The best fit parameters for this model and their 68\% errors for one
interesting parameter are summarized in Table 2; the resulting fit is
plotted in Figure 4. The best fitting values are insensitive to
whether the very central region is included in the fit or not.

According to PIMMS, an HRI count rate of 1 count s$^{-1}$ corresponds
to a flux in the 0.1 to 2.0 keV band of $3.07 \times 10^{-11}$ erg
cm$^{-2}$ s$^{-1}$ for a Raymond-Smith spectrum with a temperature of
3.4 keV absorbed by a column density of $1.4 \times 10^{20}$
cm$^{-2}$.  Using equation (5) of Henry, Briel, and Nulsen (1993)
(note that there is an error in the exponent of the initial constant
which should be $+12$ and not $-12$), we find the central electron
density to be $n_e(0) = (0.0478 \pm 0.0066)$ h$_{75}^{1/2}$ cm$^{-3}$.

\begin{table*}
\caption{X-ray properties of MS1512.4+3647 from the spatial analysis of
ROSAT HRI data.\label{tab2}}
\begin{tabular}{cccc}
\hline\hline 
                 $I_0$               &   $\theta_c$   &     $\beta$
& $\chi^2$/d.o.f.\\ 
(count ${\rm s^{-1}\;arcsec^{-2}}$)  &     arcsec     &
&                \\  
 
$(2.34\pm 0.42)\times 10^{-5}$       & $6.90\pm 1.37$ & $0.524\pm
0.031$ & $7.91/8=0.99$  \\ 
$(2.15\pm 0.40)\times 10^{-5}$       & $7.48\pm 1.54$ & $0.533\pm
0.034$ & $5.52/6=0.92$  \\ 
\hline 
\end{tabular}
\end{table*}

\begin{figure}
\vskip8cm
\special{epsfile=fig3.epsf hoffset=0 voffset=250
hscale=0.45 vscale=0.45}
\figcaption[Fig3.ps]{
Contours of the adaptively smoothed ROSAT HRI image of
MS1512.4+3647 overlaid on 1800s R band exposure obtained with the
University of Hawaii 88inch telescope in $0''.6$ seeing (from Gioia \& 
Luppino 1994, image kindly made available by I. Gioia). 
The contours are placed at constant log$_{10}$
intervals with the lowest contour being 50\% above the background.
The spacing corresponds to an increase of 25\% between adjacent
contours, resulting in 14 levels from 0.13 to 2.28 in units of
$10^{-5}$ count arcsec$^{-2}$ s$^{-1}$. \label{fig3}}
\end{figure}

\begin{figure*}
\vskip6cm
\special{epsfile=fig4left.epsf hoffset=20 voffset=150
hscale=0.4 vscale=0.4}  
\special{epsfile=fig4right.epsf hoffset=260 voffset=150
hscale=0.4 vscale=0.4}
\figcaption[Fig4.ps]{Radial surface brightness profile of
MS1512.4+3647 obtained from ROSAT HRI observations.  The best-fitting
$\beta$ model for all radial bins is superposed in the left panel. The
best-fitting $\beta$ model for all radial bins except for the central
two is superposed in the right panel. \label{fig4}}
\end{figure*}


\section{Constraints on the lensing amplification}
We are now ready to examine the lensing amplification effects on cB58
due to the cluster of galaxies MS1512.4+3647.  Our fundamental
assumptions for the construction of a model of the cluster mass
distribution are: (1) the X-ray emitting hot gas is in hydrostatic
equilibrium with the cluster's gravitational field, (2) the cluster
has a spherically symmetric mass distribution, (3) the X-ray gas is
isothermal and is in a single phase, (4) the center of the mass
distribution of the cluster coincides with the center of the cD. The
first of these assumptions is supported by the ratio of specific
energy in galaxies and gas. Using the cluster's galaxy velocity
dispersion of $\sigma=(690 \pm 96)$ km s$^{-1}$ (Carlberg et al.,
1996), we find this ratio, 
conventionally denoted by
$\beta_{\mbox{spec}}$, 
to be $0.78 \pm 0.23$, i.e.\ consistent with
unity.
The ratio is somewhat larger than $\beta$ determined by X-ray surface
brightness distribution. However Bahcall \& Lubin (1994) have pointed
out that the standard hydrostatic-isothermal model predicts
$\beta_{\mbox{spec}} = (1.25 \pm 0.1) \beta$, rather than
$\beta_{\mbox{spec}} = \beta$.
Taking this into account, $\beta_{\mbox{spec}}$ is consistent with 
$\beta$.
The fourth assumption is true based on the HRI astrometry.
Since, with ROSAT, the conversion from flux to count rate is almost
constant for hot gas in the temperature range from 2 to 10 keV, and
the temperature of MS1512.4+3647 in the cluster rest frame is much
higher than 2 keV, the gas density profile can be unambiguously
obtained from the X-ray surface brightness profile without knowing the
precise temperature distribution. According to the $\beta$ model, the
electron density distribution of MS1512.4+3647 is described by
\begin{equation}
n_e(r)=n_e(0) \left[1+\left( {r\over r_c} \right)
^2\right]^{-3\beta/2}
\end{equation}
where $r_c=\theta_c D_{OL}$ and $D_{OL}$ is the angular diameter
distance between observer and lensing cluster.  In this equation,
as in the following, $r_c/D_{OL}$ and $\beta$ are taken to be $6.90\pm
1.37$ arcsec and $0.524\pm 0.031$, respectively.

Inverting the hydrostatic equilibrium equation yields the cluster mass
contained within a radius $r$:
\begin{equation}
M(r)={kT\over G \mu m_{\rm H}} 3\beta {r^3\over r_c^2}{1\over
1+({r\over r_c})^2}
\end{equation}
where $\mu=1.3/2.1$ is the mean molecular weight for gas of cosmic
abundance. The lens equation for this mass distribution is
\begin{equation}
\tilde\theta_S= \tilde\theta_I - D {\tilde\theta_I \over
{\sqrt{\tilde\theta_I^2 + 1}}}
\end{equation}
where
\begin{equation}
D \equiv{{6 \pi \beta} \over {\theta_c}} {{kT} \over
{\mu m_H c^2}}{{D_{LS}}\over{D_{OS}}},
\end{equation}
$\theta_c$ is the angular core radius, $\tilde\theta_S$ ($
\tilde\theta_I$) is the angle between the lens and the source (image)
in units of $\theta_c$, and $D_{LS}$ ($D_{OS}$) is the angular
diameter distance between the lens (the observer) and the source 
(the explicit expression is given in Fukugita et al. 1992). 
$D$ is called the lens parameter and determines the efficiency of the
lens. It should be noted that, as the lens parameter does not depend
on the Hubble constant, neither do the amplification effects on cB58.
When $D\le 1$, the lens always produces a single image. On the other
hand, when $D > 1$, a counter image may be produced if
$\tilde\theta_S$ is sufficiently small to satisfy the condition
\begin{equation}
|\tilde\theta_S| < (D^{2/3}-1)^{3\over2}, \quad \mbox{for
$D > 1$}.
\end{equation}
The lens parameter for cB58 is $D \sim 1.4$ for $\beta = 0.52$, $k_bT
= 3.9$ keV, and $\theta_c = 6''.9$ in an Einstein-de Sitter universe,
i.e.\ $(\Omega_0, \lambda_0 ) = (1,0)$ with $\Omega_0$ being the
density parameter and $\lambda_0$ the normalized cosmological
constant. 
In the following, we shall mainly consider the following three extreme 
models; spatially flat
universes with $(\Omega_0, \lambda_0) = (1,0)$ and $(0.1,0.9)$ and the
open universe with $(\Omega_0, \lambda_0) = (0.1,0)$.

\begin{figure}
\vskip7cm
\special{epsfile=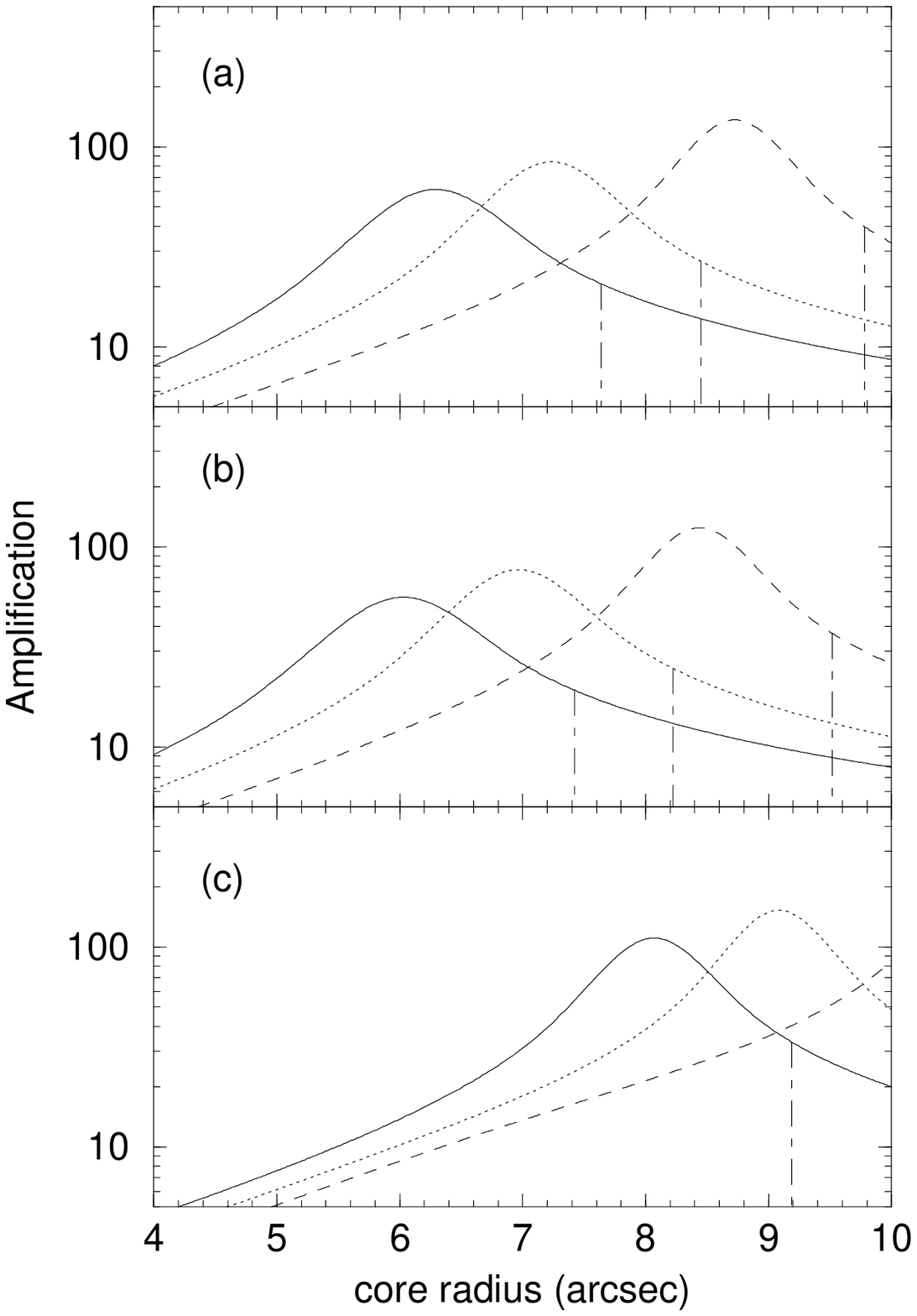 hoffset=20 voffset=250
hscale=0.4 vscale=0.4}
\figcaption[Fig5.ps]{Amplifications for $k_bT=3.6$ keV (solid lines),
3.9 keV (dotted lines) and 4.4 keV (dashed lines). The vertical lines
mark the boundary between the multiple and single image region. In the
region leftward of the line, a bright counter image appears. (a), (b)
and (c) are for $(\Omega_0, \lambda_0)=$(1, 0), (0.1, 0) and (0.1, 0.9)
respectively. \label{fig5}}
\end{figure}

Yee et al. (1996) observed cB58 with the Canada-France-Hawaii
Telescope. The seeing was $0''.64$ and $0''.73$ FWHM for the averaged
$I$ and $V$ bands respectively. With a width of $\sim 2''$
(sufficiently larger than the seeing) the image of cB58 is well
resolved.  We approximate the observed image of cB58 as an ellipse
with a semi-major axis of $1''.5$ and an axis ratio of $1.6$. See
Figure 6 and compare to Figure 2 of Yee et al. (1996).

Using the lens equation (4), we can create the unlensed source image
by mapping the observed image back onto the source plane.  The
amplification is then calculated as the ratio of the areas of the
observed and the un-lensed images.  Figure 5 demonstrates how
sensitive the amplification is to variations in the cluster core
radius and temperature.  The positions of the peaks correspond to
configurations where the observed image position coincides with a
critical line (Schneider et al. 1992).  In the region where condition
(6) is satisfied a bright counter image appears. It is important to
realize that the amplification may become large even in the single
image case and that it is possible to have a strongly magnified lensed
image without large image distortion.  The general physical mechanism
for forming such a lensed image is discussed by Futamase, Hamana
\& Hattori (1997).

\begin{figure}
\vskip5cm
\special{epsfile=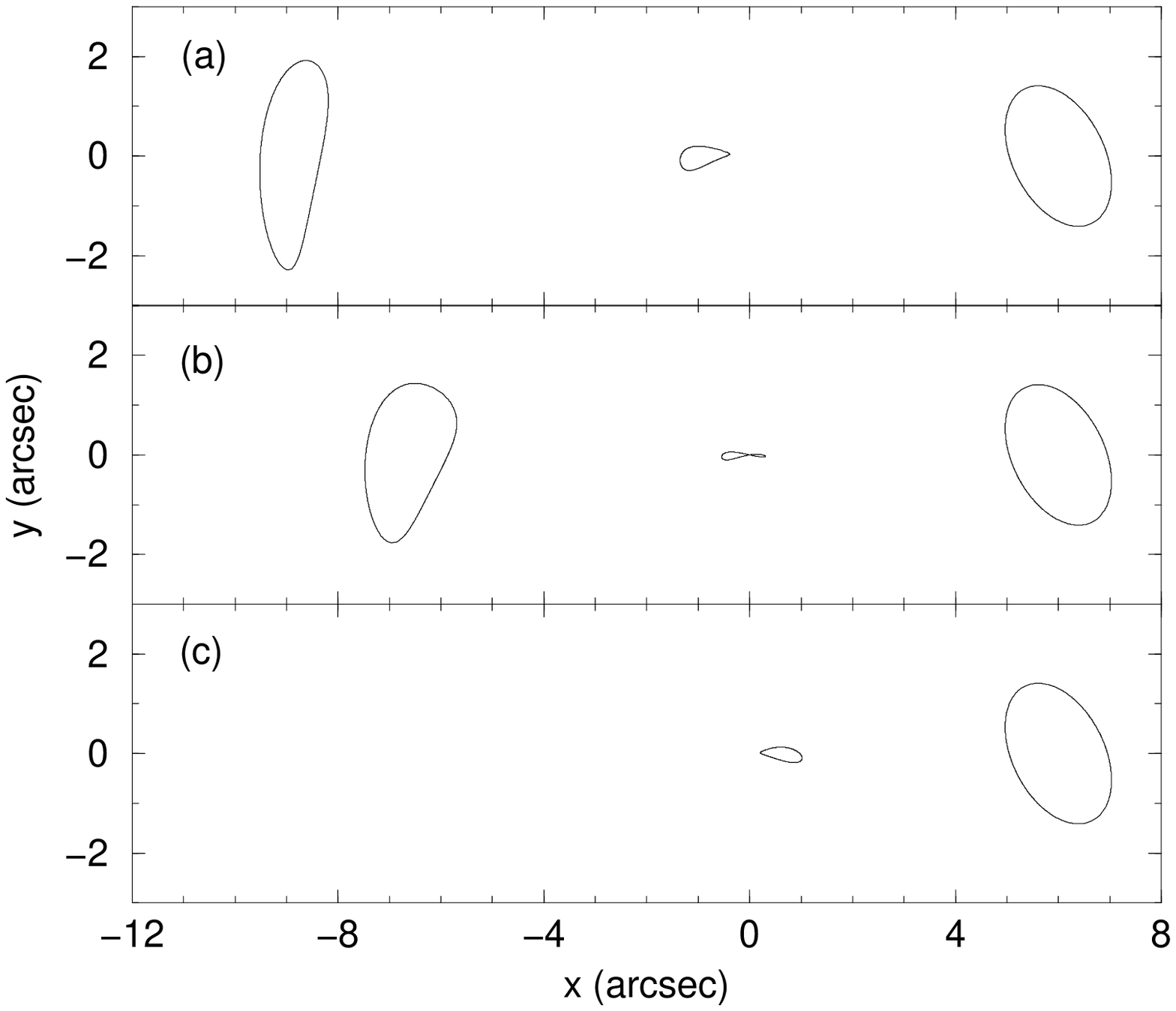 hoffset=0 voffset=220
hscale=0.4 vscale=0.4}
\figcaption[Fig6.ps]{Expected counter image (left), source (center)
and observed image (right). The plots in panels (a), (b), and (c) assume
$\theta_c=$5.5, 6.9, and 8.5 arcsec respectively; $k_bT=3.9$ 
keV and $(\Omega_0,\lambda_0)=(1, 0)$ throughout. Note that no counter
image appears in the lower panel. \label{fig6}}
\end{figure}

Figure 6 shows the counter image (left), the un-lensed source image
mapped onto the lens plane (center), and the observed image (right)
for various values of $\theta_c$.  In the case of Figure 6b, the
observed image straddles a critical line, therefore the predicted
source image consists of two parts of opposite parity. In the case of
Figure 6c, there is no counter image and the predicted source image is
less distorted.

\begin{figure}
\vskip4cm
\special{epsfile=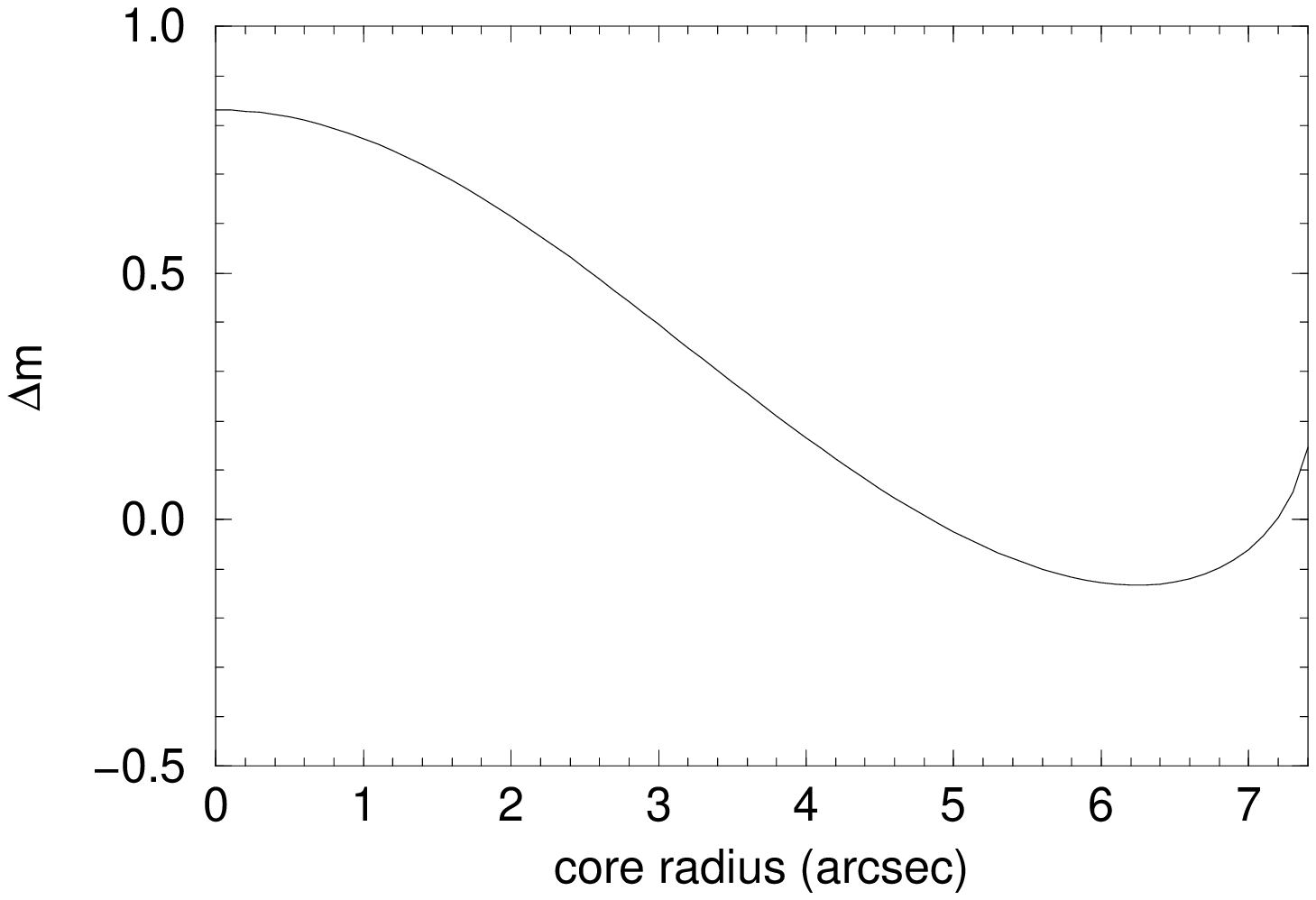 hoffset=0 voffset=220
hscale=0.4 vscale=0.4}
\figcaption[Fig7.ps]{ The difference in magnitude between the observed
and the counter image, $\Delta m=m$(observed image)$-m$(counter image)
for $k_bT=3.9$ keV and $(\Omega_0, \lambda_0)=(1, 0)$. \label{fig7}}
\end{figure}

The magnitude difference between the observed image and the counter
image is shown in Figure 7. When the core radius is small enough for a
counter image to appear, the latter must be at least as bright as the
observed image. The fact that no such bright counter image is observed
in the field around the cD galaxy, together with the measured
temperature and mass distribution for the cluster, provides strong
constraints on the amplification of the observed image.

Figure 8(a)-(c) shows the allowed region for temperature and core
radius as
well as lines of constant amplification for the single image situation
for three cosmological models, namely
$(\Omega_0,\lambda_0)=(1,0),\;(0.1,0)$, and $(0.1,0.9)$ 
respectively. From Figure 8(d), it can be
seen that a large cosmological constant reduces the single image
region drastically. 
The reason for this is the following: Condition
(6), which determines when a counter image is produced, is expressed
in terms of the lens parameter which, in turn, depends on the gas
temperature, the core radius of the cluster mass distribution and a
combination of two angular diameter distances.  The first two
quantities can be constrained by X-ray observations as shown in
Section 2.  The last one depends strongly on the value of the
cosmological constant (Fukugita, Futamase \& Kasai 1990).  Therefore,
the value of the cosmological constant can, in principle, be
constrained by the requirement that there be no counter image of cB58.
This new test was first proposed by Futamase, Hamana \& Hattori
(1997), and cB58 is the first and very good application of this
method.

\begin{figure}
\vskip8cm
\special{epsfile=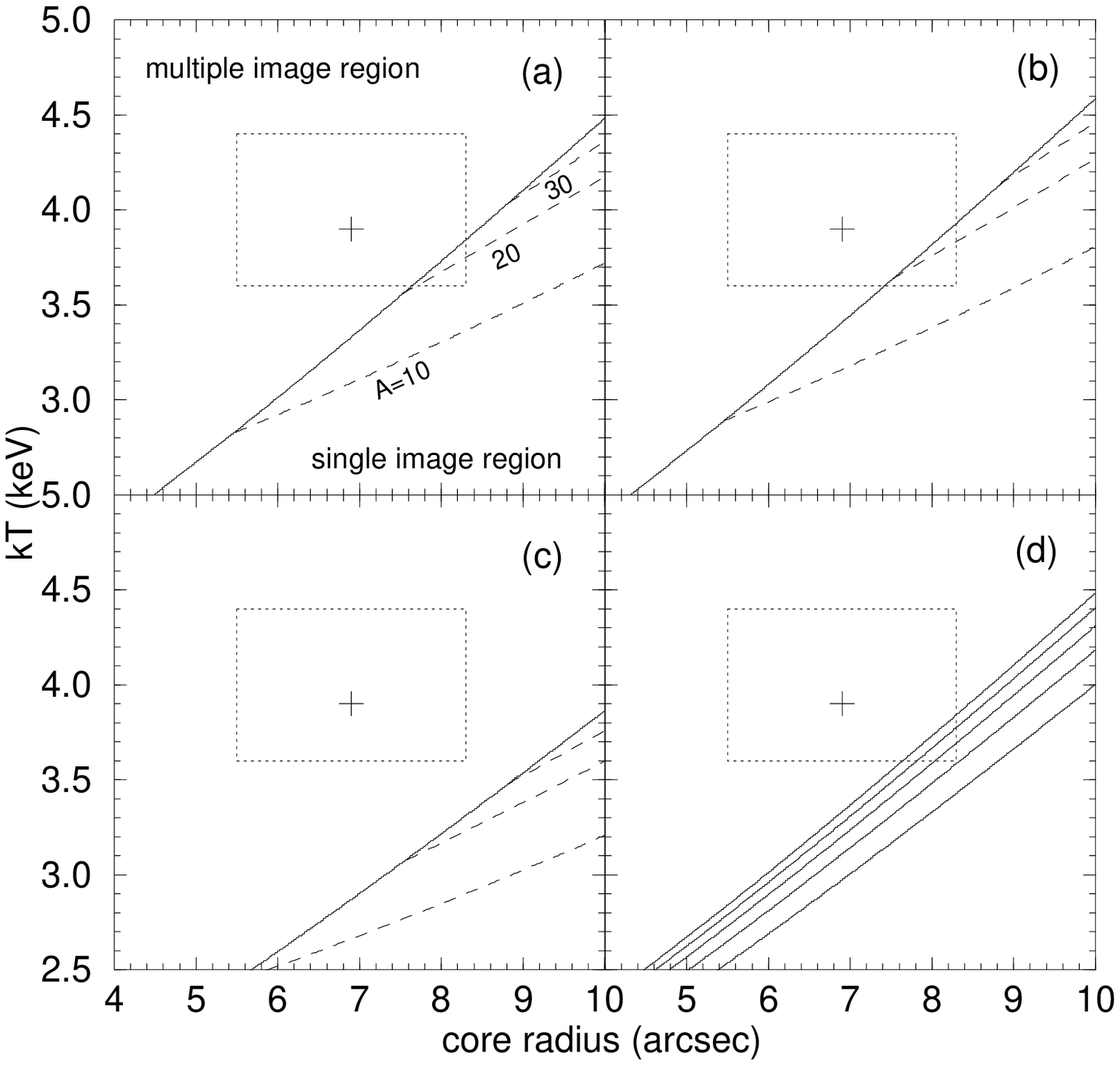 hoffset=-20 voffset=320
hscale=0.5 vscale=0.5}
\figcaption[Fig8.ps]{Allowed regions in the $\theta_c-k_bT$ plane. The
solid line marks the boundary between the multiple and single image
region. A counter image is expected for $\theta_c-k_bT$ combinations
that lie above and left of the solid line. Lines of constant
amplification are shown dashed.  The plus and dotted box denote the
best fit and 68\% errors of the core radius obtained from ROSAT HRI
data (for all radial bins) and the X-ray temperature obtained from
ASCA and ROSAT PSPC data, respectively.  (a), (b) and (c) are for
$(\Omega_0, \lambda_0)=$(1,0), (0.1, 0), and (0.1, 0.9) respectively.
(d) The boundaries are shown for universe models with $\lambda_0 +
\Omega_0 =1$. From upper to lower $\lambda_0=0, 0.2, 0.4, 0.6 $ and
$0.8$.
\label{fig8} }
\end{figure}

If the best fit value of the temperature from the ASCA and ROSAT PSPC
analysis and the best fit value of the core radius from the ROSAT HRI
analysis are assumed, any universe model with reasonable cosmological
parameters predicts a bright counter image of cB58 which is not
observed.  If the $1\sigma$ errors in temperature and core radius are
taken into account, both an open universe and a flat universe with
zero cosmological constant are acceptable and predict that the
amplification of cB58 can be as large as 25. However, only a small
part of the temperature--core radius region is allowed.  On the other
hand, a flat universe with a large cosmological constant does not have
any acceptable region.

However, if $2\sigma$ or $3\sigma$ measurement errors are taken into
account, the above mentioned constraints are relaxed.  The statistical
errors in the temperature is large: $k_bT=3.9^{+0.9}_{-0.6}(2\sigma)$,
${}^{+1.5}_{-0.9}(3\sigma)$.  A flat universe with a large
cosmological constant can have an acceptable region.  A large range of
amplifications from several to 25 is possible regardless of the
cosmological models.  If the lensing amplification of cB58 were 50,
its intrinsic star formation rate would be reduced to a value found
for other distant galaxies (Steidel et al., 1996a). Hence, although
such a high amplification is improbable, we can not definitely
determine whether cB58 really features a very high star formation rate
or is just a normal starburst galaxy at high redshift.

How sensitive is this result to changes in our model assumptions? One
of our initial assumptions was that the X-ray emitting gas is
isothermal and in a single phase. If a cooling flow existed in
MS1512.4+3647 it would cause the gas to be multiphase in the central
region of the cluster.  Allen et al. (1996) claimed that for clusters
hosting strong cooling flows the use of a single phase model can lead
to significant underestimates of the overall temperature. Accordingly,
a strong cooling flow in MS1512.4+3647 would mean a higher lower limit
to the lensing amplification and more severe constraints on the
cosmological constant.  However, contrary to the results from earlier
work (Stocke et al, 1991), we find the evidence for a cooling flow to
be marginal for this cluster.

We also assumed that the cluster has a spherically symmetric mass
distribution. The ROSAT HRI image (Figure 3), however, shows the X-ray
emission to be assymmetric outside of the core region.  Bartelmann
(1995) has pointed out that deviations from spherical symmetry in
cluster lenses enhance the tidal effect (shear) and alter the lensing
efficiency of the cluster.  However, to allow the effects of the
asymmetry in MS1512.4+3647 to be quantified, an accurate spatial model
of the mass distribution is required. Such a model is now being
developed.

The observations presented here may be used to measure the gas mass
fraction in this cluster. Inserting the central gas density found in
Section 2.2 into equation (2) and integrating out to the maximum
radius detected ($100''$ or 0.42 h$_{75}^{-1}$ Mpc), we find a gas
mass of $(2.14 \pm 1.74) \times 10^{13}$ h$_{75}^{-5/2}$
M$_{\odot}$. This mass is proportional to the core radius cubed, which
accounts for the large error, see Table 2. The total mass out to the
same radius, from equation (3) is $(9.20 \pm 1.09) \times 10^{13}$
h$_{75}^{-1}$ M$_{\odot}$. Hence the gas mass fraction is $(0.23 \pm
0.19)$ h$_{75}^{-3/2}$. This value, although poorly constrained, is
consistent with that of low redshift clusters (White and Fabian,
1995).
 
\section{Conclusion}

We have examined the amplification of cB58 due to gravitational
lensing by the foreground cluster of galaxies MS1512.4+3647 based on
ROSAT and ASCA X-ray observations of the cluster.  Unfortunately, with
the available X-ray observations we cannot specify the physical
parameters of the foreground cluster MS1512.4+3647 accurately enough
to evaluate the lensing amplification and definitely determine whether
cB58 is a normal star forming galaxy.  We have also shown that cB58
allows a new and promising test to constrain the value of the
cosmological constant.  
In this study, however, we cannot give a strong constraint on the
cosmological constant, because of large statistical errors in our X-ray
observations. 
If one could measure
an accurate mass distribution for MS1512.4+3647, particularly its
ellipticity, then the observational fact that no counter image of cB58
is observed would give a strong constraint on the cosmological constant.
From both points of view, it is highly
desirable to perform further observations to obtain more accurate
information on the temperature and the mass distribution of the
cluster MS1512.4+3647.

\acknowledgements

It is a pleasure to thank T. Yamada for useful discussions. We also
thank I. Gioia for making available to us the optical image of
MS1512.4+3647.
MH thanks
D. Neumann and T. Miyaji for useful comments on X-ray data analysis.
MH would like to thank the Max-Planck-Institut and the Yamada Science
Foundation for financial support. HE gratefully acknowledges financial
support from SAO contract SV4-64008, JPH's research has been supported
by NASA grants NAG\,5-1880 (ROSAT) and NAG\,5-2513 (ASCA) and by NSF
grant AST\,95-00515.  This work is supported in part by Japanese
Grant-in-Aid for Science Research of Ministry of Education, Science
and Culture, No. 07640366.

\end{document}